\documentclass[11pt,aps,tightenlines,nofootinbib,longbibliography,superscriptaddress]{revtex4-1}
\usepackage{graphicx}
\usepackage{amssymb}
\usepackage{amsmath}
\usepackage[margin=1.25in]{geometry}
\usepackage[usenames,dvipsnames]{color}
\usepackage{url}
\usepackage[colorlinks = true,
            linkcolor = blue,
            urlcolor  = blue,
            citecolor = blue,
            anchorcolor = blue]{hyperref}

\newcommand\snowmass{\begin{center}\rule[-0.2in]{\hsize}{0.01in}\\\rule{\hsize}{0.01in}\\
\vskip 0.1in Submitted to the  Proceedings of the US Community Study\\ 
on the Future of Particle Physics (Snowmass 2021)\\ 
\rule{\hsize}{0.01in}\\\rule[+0.2in]{\hsize}{0.01in} \end{center}}

\begin{document}

\title{Snowmass 2021: Superconducting Sensor Fabrication Capabilities for HEP Science}

\author{Thomas Cecil}
\affiliation{Argonne National Laboratory, Lemont, IL}
\author{Clarence L. Chang}
\affiliation{Argonne National Laboratory, Lemont, IL}
\author{Shannon M. Duff}
\affiliation{National Institute of Standards \& Technology, Boulder, CO}
\author{Dale Li}
\affiliation{SLAC National Accelerator Laboratory, Menlo Park, CA}
\author{Rupak Mahapatra}
\affiliation{Texas A\&M University, College Station, TX}
\author{Mark Platt}
\affiliation{Texas A\&M University, College Station, TX}
\author{Kevin K. Ryu}
\affiliation{MIT Lincoln Laboratory, Lexington, MA}


\snowmass

\maketitle


\section{Executive Summary}
Superconducting sensors are a key enabling technology for many HEP experiments with advances in sensor capabilities leading directly to expanded science reach. The unique materials and processes required for the fabrication of these sensors makes commercial sourcing impractical 
in comparison with semiconducting devices. Subsequently, the development and fabrication of new sensors are often performed at academic cleanrooms supported through HEP basic detector research and/or project funds. While this operational model has been successful to date, we are at a turning point in the history of superconducting electronics, as evidenced by the rapid growth in the field of quantum computing, when scale and sophistication of these sensors can lead to significant progress. In order to achieve this progress and meet the needs of the next generations of HEP experiments, continued support of all stages of the superconducting sensors development pipeline is necessary.

\section{Introduction}

Superconducting sensors have a rich history in High Energy Physics (HEP). For example, Transition Edge Sensors (TESs)\cite{Irwin2005} have been deployed in WIMP Dark Matter searches \cite{Brink2009} (using hundreds of TESs on a large crystal in order to perform event reconstruction from athermal phonons) and in measurements of the Cosmic Microwave Background (CMB) (e.g., the SPT-3G experiment \cite{Anderson2018} utilizes a focal plane of over 10,000 TES bolometers to search for signs of inflation in the polarization of the CMB). Another HEP application of superconducting circuit technology is the ADMX experiment \cite{Du2018} which uses a superconducting quantum interference device (SQUID) to amplify the potential microwave signal produced from axion Dark Matter interacting with a magnetic field.
Because the typical energy scale for superconductivity is the gap energy, $\mathcal{O}(10^{-4})$~eV, detectors and devices that exploit superconductivity are capable of measurements that are difficult using other technologies such as semiconducting devices, which have characteristic energy scales of $\mathcal{O}(1)$~eV. Superconducting sensors are especially well suited to applications where the signal is rare or faint, making sensitivity critical. 

The many unique properties of superconductors, such as zero or near zero resistance to signals up to microwave frequencies and non-linear kinetic inductance, enable a wide range of detector technologies and applications: the TES is a superconducting thermistor that can be used in both bolometeric and calorimetric modes and used for photon detection covering frequency ranges from mm-wave to gamma ray; the kinetic inductance detector (KID)\cite{mazin2009microwave} is a superconducting resonator with a surface inductance sensitive to quasiparticle density that is inherently multiplexible in the frequency domain; the superconducting nanowire single photon detector (SNSPD)\cite{snspdreview} is a thin, narrow superconducting strip biased near its critical current that can sense single photons with high rate, lower jitter ($\sim$100~ps) and very low dark count; the SQUID\cite{squidhandbook} has been used as a sensitive magnetometer, ammeter, and tool for multiplexing other cryogenic detectors; the quantum bit, or qubit, which serves as the fundamental building block of quantum computers, can also be used as a sensitive photon detector \cite{Dixit2020}; and the parametric amplifier enables amplification of signals with noise levels approaching the quantum limit \cite{HoEom2012}. 

 Despite the wide use of superconducting sensors, there are still several challenges associated with fabricating these complex integrated devices while maintaining the required control of the underlying materials properties. Future technical goals include realizing lower detection thresholds, larger collecting areas, increased sensor count, and increased integration of detector components.
 Addressing these challenges will require a sustained effort at all levels of the detector development life cycle - R\&D, prototyping, and production - and an increased focus on collaboration between the many facilities currently working in this area. Below we highlight some of the key features that distinguish the superconducting fabrication process, discuss existing capabilities, and consider future needs to continue advancing the field of superconducting sensors for the benefit of HEP science.

\section{HEP Science with Superconducting Sensors}    

HEP experiments utilizing superconducting detectors require the custom creation of complex detectors integrated into specific experiment configurations which is justified given the measurement capabilities these detectors provide. Looking towards the future, superconducting detectors, devices and circuits will continue to play a critical role across a number of HEP science thrusts. Potential examples, though by no means an exhaustive list, include:
\begin{itemize}
	\item {\bf Studying early and late cosmic acceleration via mm-wave cosmological surveys:} Upcoming and future mm-wave cosmological surveys are powerful tools for understanding new physics associated with the cosmic acceleration of the universe~\cite{Snowmass2021:CMBexp,Snowmass2021:mmLIM}. At the largest spatial scales, these surveys can explore multiple signals associated with the physics of inflation, with CMB experiments being unique in their potential to measure the inflationary energy scale. At smaller spatial scales, these experiments will precisely measure the evolution of our universe’s energy content providing stringent constraints on the physics of the dark sector, dark energy, and neutrinos. For these surveys, the individual detector sensitivity is already fundamentally limited by the equivalent of photon shot noise. Thus, the critical technical challenge for these mm-wave surveys is the successful fabrication of a large number of high sensitivity mm-wave detectors. Upcoming and future experiments~\cite{Snowmass2021:CMB-S4, CMB-HD-Snowmass,Snowmass2021:mmLIM} envision detector payloads with 500,000 to over a million detectors. Superconducting detectors along with their associated superconducting electronics are the only technology to demonstrate the required combination of sensitivity and scalability, and realizing these future instruments will need facilities-scale research, development and support.
	\item {\bf Searches for low-mass Dark Matter particles and CE$\nu$NS:} Searches for low-mass Dark Matter strongly benefit from the ability to reach as low a threshold as possible to take advantage of the featureless exponential recoil energy spectrum. Searches for Coherent Elastic Neutrino Nucleus Scattering (CENNS) interaction also relies on very low-threshold detectors, especially in searches for new physics with reactor anti-neutrinos. Both of these research areas aim to reach thresholds as low as few eV, but definitely sub-keV. 
	
	Reaching such low-threshold is usually realized through detection of phonons through superconducting sensors, such as: TESs used in the Super Cryogenic Dark Matter Search (SuperCDMS) experiment to achieve eV-scale recoil energy resolution~\cite{Brink2009,Snowmass2021:CDMS}; exploring SNSPDs with sub-eV threshold for direct detection of sub-GeV Dark Matter \cite{Hochberg2019}; and developing low-threshold TES detectors and KIDs~\cite{DMoore2012,YYChang2018} for calorimetric measurements of photons, light, heat, and evaporated Helium atoms for light Dark Matter particle searches \cite{Hertel,Derenzo2017,Knapen2018}. Low-threshold TES detectors are being developed for use as light detectors for background discrimination in neutrino-less double beta decay searches \cite{TheCUPIDInterestGroup2015} and as thermistors for calorimetric measurement of Coherent Elastic Neutrino Nucleus Scattering using dielectric or superconducting crystals \cite{Billard2017}.
	
	Phonon detectors usually avoid the strong suppression of signal from nuclear recoil by not relying on the dissipation channels that are suppressed due to the Lindhard effect, such as the ionization production in Si CCDs or Ge PPC detectors. In these phonon-assisted detectors the usual technology is to use superconducting sensors made of material that have low superconducting transition temperatures and have superconducting bandgap energies (typically few meV) that are lower than the highest phonon energies. 
	\item {\bf Searches for Axion and Axion-like Dark Matter:} Axions and Axion-like particles are an attractive class of dark matter candidates  where the dark matter consists of very light particles. Within this class of candidates the QCD axion would provide a solution to the strong-CP problem and more general ALPs appear in multiple extensions of the Standard Model. Searches for this class of dark matter candidates typically involve measuring a very faint low frequency ($\lesssim$ 10 THz) photon signal arising from dark matter-to-photon conversion in the presence of a strong magnetic field. Superconducting technologies have world leading sensitivity in these searches (see~\cite{PDG2020}) and have a central role in a number of proposed quantum techniques aimed at achieving sensitivity beyond the standard quantum limit~(e.g. \cite{Dixit2020, QuantizedAwards}).
\end{itemize}


\section{Superconducting Sensor Fabrication Processes}

The wide range of superconducting sensors previously mentioned is bound by a set of common fabrication processes. These processes have been developed and improved upon over many generations of superconducting devices, yet lag behind progress in semiconductor microfabrication due to funding differences and the lack of sufficient fabrication facilities and toolsets to advance superconducting detector fabrication capability to its full potential. Despite this, the advancement of fabrication processes for superconducting integrated circuits has continued to progress and materials like low-loss dielectrics or engineered superconductors (bi-layers) have been continually refined. The fruits of this effort can be seen in the deployment of advanced sensors in current experiments (e.g., 3rd generation CMB, CDMS, and ADMX) and in the recent success of quantum computing. There is an opportunity now to open the field to even larger success – to realize the successes demonstrated in the scale and sophistication of quantum computing circuitry to HEP sensors.

There are many examples of important advancements in superconducting fabrication processes in recent years – too many to cite them all.  One such example is the multi-layer niobium process at MIT/LL, a demonstration of the type of process development required in superconducting device fabrication to achieve the next level of complexity and competitiveness with CMOS technology \cite{Tolpygo2015, Tolpygo2016, Tolpygo2019}.  Achieving this milestone required process development in all areas: photolithography, film deposition uniformity and repeatability, chemical mechanical polishing for planarization, via etch processes, metal etch processes, and low-loss dielectric deposition. Similarly, many 2nd to 3rd generation CMB experiments developed fabrication processes to enable the superconducting detector arrays to be made on 150~mm substrates (previously fabricated on 3~inch substrates) \cite{Duff2016, Posada2015}. This allowed for a higher detector filling fraction and, therefore, a higher sensitivity instrument while reducing the number of sensor wafers and time to fabricate. Development of sensor materials with reliable critical temperatures across a larger format wafer was a large part of this success \cite{Li2016}.

It is worth noting that many superconducting devices require the use of “exotic” materials, like gold or aluminum, which may be incompatible with typical semiconductor processes. Availability of these materials and robust processes utilizing these materials is critical for required device performance. In addition to niobium, a superconducting material often used for integrated circuit wiring in superconducting electronics, integration of these exotic materials for the sensor and low-loss dielectrics make up a large fraction of the required processes for successfully yielding superconducting detectors. All of these films require a high level of process control in order to repeatably yield the same device over and over. Moreover, deposition of these specialty materials often requires the ability to tune the film stress during deposition as well as maintain an ambient temperature at the substrate, features not required in other types of fabrication. Contamination issues commonly arise in systems exposed to magnetic materials (which spoil superconductivity) or materials with any radioactive impurities (which spoil sensitivity of sensors and detectors).


In addition to tight process controls, successfull fabrication of supercoducting sensors require a tightly controlled environment with regulation of humidity, temperature, particle count, laminar air flow, and ultra-pure water to fabricate pristine devices with multiple layers of complexity. Ideally, this happens in a dedicated clean facility (ISO 5/Class 100 or better) with all tools housed within the same clean space. In addition to the unique material needs, superconducting fabrication requires the integration of many specialized techniques including but not limited to micro-electro-mechanical systems (MEMS) in the form of thin membranes and high aspect-ratio cantilevered legs, small area Josephson Junctions, high uniformity for monolithic arrays of detectors, and full wafer wiring.

To support the fabrication of superconducting sensors, a comprehensive suite of metrology tools are used to characterize, diagnose, and troubleshoot fabrication process development.  These metrology tools include optical microscopy, ellipsometry, spectrophotometry, profilometry, dc/ac probing, film stress measurement, surface roughness measurement, scanning electron microscopy, and focused ion beam microscopy. In addition, superconducting fabrication requires the use of cryogenic systems outside the cleanroom for rapid feedback of performance characteristics including critical temperature, residual resistance ratio, critical current, subgap potential, nonlinear inductance participation, microwave loss tangent, resonant quality factor, thermal conductance, heat capacitance, etc. Understanding the correlation between these performance characteristics and the fabrication parameters are crucial to the science of superconducting sensor fabrication.
    

\section{Existing capabilities}
The current ecosystem of superconducting sensor fabrication facilities occurs at a wide range of locations including universities, national laboratories, federally funded research and development centers, and companies. These facilities have capabilities covering the full range of detector development from basic materials research to device prototypes to fully functioning detectors on scales of a few wafers to a few hundred wafers. Often a single physical cleanroom facility will host equipment for several phases of the development pipeline. In the research and development phase the focus is on new materials and devices and requires equipment that can be quickly adapted to new processes but can often result in a lack of stability. The general purpose nature of this equipment often makes it difficult to handle the specialized substrates that are needed for many HEP sensors requiring the purchase and maintenance of custom systems. Moving into the prototyping stage requires an increased level of process control that often precludes sharing of process resources with those used for research, or at a minimum requires blocking off equipment time. For full scale detector production dedicated processing equipment is required, both for the processes being run and the time the equipment is available for a given project. 
Lacking the large scale commercial market of semiconducting devices, much of the development capabilities are supported through field work proposals or individual projects. Facilities have leveraged this support to advance new detector concepts for the next stage of experiments, but as new experiments grow in scale and complexity this model will no longer be sufficient. Critically, broad support is needed for all the phases of detector development.

One of the most important capabilities that is often overlooked is a well-trained workforce. Dedicated staffing with long-term process knowledge is needed as the complexity of these sensors increases. Currently the main drivers of sensor development are often new students. As they begin working on a project they will devote considerable time to fabricating sensors, only to transition to data taking and data analysis once a prototype has been fabricated, leaving the cleanroom and taking their process knowledge with them. As projects become more complex the development times will extend past the length of a single student or postdoc. Additionally, with the growth of quantum computing the demand for workers with skills in superconducting sensor fabrication will increase, diverting these workers away from the HEP community. A robust program for workforce training is required in addition to clear pathways to applying these skills in careers developing HEP sensors. 

\section{Future needs}
Looking forward over the next decade and beyond there are many exciting experiments that will rely on superconducting detectors. Achieving the requirements of these detectors will require advances in several areas including new materials, detector mounting and interconnect methods, fabrication facilities, and methods of coordination in the community. Below we highlight a few of the key needs of each of these areas and the impact that can be had on future sensors.
\subsection{New Materials}   
Improvements in sensor performance often start with new materials pushing the boundaries of what is possible for a given sensor design. One of the advantages superconducting detectors offer relative to other detector technologies is a low energy threshold due to the small pairing energy of cooper pairs (O(meV) compared to O(eV) for semiconductors). The energy required to split a cooper pair is directly related to the transition temperature of the superconductor (T$_C$). Thus, developing detectors with lower thresholds for applications such as direct dark matter searches 
or CEvNS requires finding new material systems with lower transition temperatures. Several possible methods for engineering the transition temperature exist including bi-layers, doping or compound materials, and phase selection of pure materials.  As the transition temperature, and operating temperature of sensor, approaches the range of a few mK understanding the fundamental physics of heat transfer in the material system becomes increasingly important.

Future spectroscopic survey techniques such as Line Intensity Mapping (LIM) will require superconducting filter banks with increased resolution of order R = 300 \cite{Snowmass2021:mmLIM}. The resolution of these spectrometers is often dominated by the loss of the dielectric used in transmission lines and channelizing filters. New dielectric materials with lower losses at mm-wave frequencies are required. Promising candidates have been found such as low-loss SiN, a-Si and a-SiC, but a significant breakthrough is limited by a lack of understanding of the fundamental processes determining the loss at these frequencies. The need for low-loss dielectrics extends to any device that utilizes on-wafer wave-guides including traveling wave parametric amplifiers (TWPAs) and qubits. 
Many different devices including MKIDs and TWPAs take advantage of the kinetic inductance of superconductors. In MKIDs this drives the frequency shift of the resonators as a function of quasi-particle density and in TWPAs the non-linear nature of the inductance is means for wave mixing. New materials with higher kinetic inductance would enable KIDs with increased sensitivity and TWPAs that can be operated with smaller pump powers.

 \subsection{New Detector Mounting and Interconnects}
 Development of detector mounting and wiring technologies is another critical area as the detector sensitivity increases. The hardware that was used to mount detectors in the past has now proven to be a source of background noise. Common mounting materials such as Cirlex are failing to meet the radio pure needs of new detectors. The challenge of getting the signal from the surface of the detector to the outside world is an ongoing struggle that grows in complexity as detector performance increases. Detector innovation drives improvement of everything that touches the detectors.
 
 3-D integration is needed for quantum circuits, integrated systems of detectors and readout electronics, and 'cavity' resonators at a smaller scale. Many cryogenic detector systems require detector, amplification, and multiplexing elements. These components are often fabricated on separate wafers due to varying fabrication requirements for each circuit and in order to limit the risk of a single component failure causing total system failure. However, as these systems scale integration of the components becomes increasingly complex and can result in signal loss or interference due to multiple stages of interconnect wiring. In some applications, impedance matching dictates short lengths of wiring to limit parasitic inductance (e.g., rf-mux). 3-D integration of multiple components via wafer stacking and bump bonding or other technologies can reduce module complexity and, in some cases, lead to higher-density detector arrays and circuits. Initial demonstrations of 3-D integration of quantum circuits have been completed\cite{yost2020solid}, but further development of the process toward the desired complexity is needed. 
 
 \subsection{New Fabrication Facilities}
 As the field delves into different substrate materials, shapes and circuit architectures to reduce noise and increase signal, the device fabrication becomes more complex. The standard retro tool set for semiconductor circuits does not work for all facets of superconducting sensor fabrication. Thankfully the maturity of the semiconductor industry has provided a large variety of equipment and techniques that can be adapted with equipment and processes from other industries to meet the needs of HEP. Thinking outside of the wafer is required to achieve our goals but comes with an additional cost.
 
 Increasing resolution of circuit fabrication calls for a more up-to-date tool set. The jump from micron resolution to sub-micron resolution requires a host of newer technologies such as E-Beam Lithography (key for patterning SNSPDs) and Deep Reactive Ion Etch (key for optical coupling structures for CMB sensors). E-Beam Lithography is a maskless lithography process used to pattern the layers of the device circuit. Being maskless lends itself well to the research and development environment as the pattern is programmed rather than made into optical mask plates. This pattern program can be adjusted substrate to substrate based on test results. This tight closed loop can accelerate research progress dramatically. Deep Reactive Ion Etch is a highly directional plasma etch that can be used to form the sub-micron structures. In tandem technologies such as these and many other newer technologies can improve device resolution, decrease noise and increase sensitivity. 
 
 Developing new equipment and techniques for substrate shaping and smoothing can open possibilities in the bulk structures of future detectors. As the physics evolves away from standard semiconductor shapes, such as wafers and ingots, the surface preparation becomes more complex. Fabrication on non-standard semiconductor shapes such as creating field shaping lines on the sides of detectors also push the envelope. The physics should dictate detector design not the legacy fabrication hardware handed down from yesterdays commercial fabrication facility. Both the Dark Matter and the CENNS experiments require large (kg-100 kg) payloads with low-thresholds to achieve sufficient sensitivities for new physics. The ability to fabricate superconducting sensors on large bulk material (semiconductors, scintillators) is critical to the future of these physics searches.
 
 Pushing further, an underground fabrication facility allows for the fabrication of detectors with very low cosmogenic activation. Space in under ground clean room facilities such as SNOLab can be converted into detector manufacturing areas. It is possible to create a facility in which a crystal boule can be pulled, aligned, shaped, polished, and ingots fabricated into detectors deep underground. If said facility was shared with the corresponding experiment a detector could be placed in service with out ever being exposed to the cosmic bombardment of the Earth's surface.

\subsection{Coordination}
Many organizations have specialized fabrication capabilities that are unique to that organization. In this environment, collaboration offers a path to leap frog detector capabilities while keeping risk, schedule, and cost down. In order for the collaboration to be successful, it requires a level of trust between the two organizations so that technical details can be shared and a level of caring for the common purpose of the project. In some cases this could involve the sharing of intellectual property as has happened recently in the sharing of design layouts in the CMB-S4 collaboration. These interactions require buy-in from organizational management and creation of memorandum of understanding, or other vehicles for protecting each institution.

In other cases this coordination can take the form of a split fabrication process. Recent demonstration of a 100 kilo-pixel microcalorimeter array for x-ray imaging spectroscopy (XIS) from NASA Goddard Space Flight Center (GSFC) is a good example \cite{devasia2021large}. Cryogenic microcalorimeters have demonstrated x-ray energy resolution that is 60x better than silicon CCDs. These detectors offer single photon sensitivity with very low dark noise and ability to provide both imaging and spectral discrimination of E/$\Delta$E of greater than 3000 for 6~keV x-rays \cite{bandler2016development}. However, compared to CCDs, they lack number of pixels and field of view. The largest microcalorimeter array that has been flown is only 128 pixels \cite{adams2020first} while CCDs on board Chandra X-ray Observatory have 1 megapixel each \cite{burke1997soft}. 

In 2017, one of the main challenges for scaling up the microcalorimeter array size at NASA/GSFC was the challenge of building high-density and high-yield microstrip superconducting wiring on the microcalorimeter wafer. Each sensor in the array needs a pair of superconducting wires to bring the signal in and out of the sensor from the center of the array to the edge, where it can be connected to a read out chip. As the array size grew, the number of wires required to squeeze between the sensors increased and that limited the size of the array that could be built. At MIT Lincoln Laboratory (MIT/LL), a process to integrate eight layers of sub-micron superconducting Nb wiring had been developed to demonstrate high density superconducting single-flux quantum (SFQ) circuits using Josephson junctions \cite{tolpygo2014inductance}.

NASA/GSFC formulated a collaboration to fabricate the superconducting wiring at MIT/LL and deliver the wafers to NASA/GSFC to add on the microcalorimeters. 
With support from both organizations, a risk reduction plan utilizing internal research and development funding started soon after to address risks related to different wafer sizes and material interfaces between fabrication sites. 
Prior to starting a fabrication  run, design rules and process design kits (PDK) were shared to ensure that no design rule is violated and that keep out zones for chemical mechanical polishing fill patterns are properly defined. The collaboration started in the summer and by the end of the year, a demonstration of a high performance microcalorimeter array was complete reducing risk for this path. 
As a result of the collaboration, NASA/GSFC was able to fabricate a 100 kilo-pixel microcalorimeter array which would have taken longer and cost more to develop without the collaboration. 

This is just one example and while not all detectors may benefit from collaboration, it is possible a large number of HEP detectors will. In order for collaboration to be successful, small risk reduction experiments should be done fast, while increasing the chance of success by setting up multiple ways the experiments can succeed. 
Collaboration is a strategy that is consistent with past HEP detectors, where commercial detectors are employed in experiments. In a low volume fabrication of detectors, it enables leveraging of key technology that was developed for other projects in support of HEP detectors. If employed properly, it will enable more capable detectors at a given cost leading to more scientific discoveries. 




\section{Summary}
Superconducting sensors provide energy sensitivity that is challenging to achieve using traditional sensors. They are already contributing in many high energy physics experiments to study CMB, Dark Matter, and neutrinos. With the recent investment in superconductor quantum computing there is now an opportunity to increase the scale and integration of these sensors for even greater impact. In order to realize this opportunity, investments should be made in the fabrication of these sensors as superconducting sensors require unique set of materials and processes not typically supported by commercial CMOS foundries. The current ecosystem of superconducting fabrication facilities occurs at universities, national laboratories, FFRDCs, and companies. Broad support is needed for all phases of sensor development - from material development to full scale production dedicated for large experiments. This will enable the pipeline of new sensors with better performance while performing practical field test experiments and training work force in superconducting sensors. 

Material development such as lower T$_C$ superconductors and lower loss dielectrics can benefit future sensors. In addition, new packaging solutions with higher radio purity and 3D integration with the ability to integrate large numbers of sensors is desired. For sensors requiring high radio purity, an underground fabrication facility can produce sensors that have not been exposed to the cosmic ray bombardment on Earth’s surface. For sensors where sufficient volume is not available to justify the cost of a dedicated facility coordination between organization offers an opportunity to produce high quality sensors at reduced schedule and cost.

The recent explosion of interest in superconducting devices via quantum computers provides an unique opportunity to design large scale experiments utilizing superconducting sensors. A strategic investment in key areas will enable leveraging of existing efforts to train the future work force in this field and produce sensors for exciting HEP experiments.

\section{Acknowledgments}
MIT Lincoln Laboratory portion of this material is based upon work supported by the National Aeronautics and Space Administration and Under Secretary of Defense for Research and Engineering under Air Force Contract No. FA8702-15-D-0001. Any opinions, findings, conclusions or recommendations expressed in this material are those of the author(s) and do not necessarily reflect the views of the National Aeronautics and Space Administration and Under Secretary of Defense for Research and Engineering.





\bibliographystyle{JHEP}
\bibliography{references.bib}  










\end{document}